\documentclass{article}

\usepackage{arxiv}

\usepackage[utf8]{inputenc} 
\usepackage[T1]{fontenc}    
\usepackage{hyperref}
\hypersetup{
    colorlinks=true,    
    citecolor=blue,
    linkcolor=red,
    urlcolor=blue,
    pdfborder={0 0 0}    
}     
\usepackage{url}            
\usepackage{booktabs}       
\usepackage{amsfonts}       
\usepackage{nicefrac}       
\usepackage{microtype}     
\usepackage{lipsum}
\usepackage{graphicx}
\graphicspath{ {./images/} }
\usepackage{amsmath}
\usepackage{comment}
\usepackage{setspace}
\usepackage{cite}

\title{Computational Fluid Dynamics: its Carbon Footprint and Role in Carbon Emission Reduction}

\author{
Xiang I A Yang \\
  Mechanical Engieering\\
  Pennsylvania State University\\
  State College, PA 16803 \\
  \texttt{xzy48@psu.edu} \\
   \And
 Wen Zhang \\
  Mechanics and Aerospace Engineering\\
  Southern University of Science and Technology\\
  Shenzhen 518055, China \\
  \texttt{zhangw6@sustech.edu.cn} \\
  \And
 Mahdi Abkar \\
  Mechanical and Production Engineering\\
  Aarhus University\\
  Aarhus N 8200, Denmark \\
  \texttt{abkar@mpe.au.dk} \\
  \And
 William Anderson \\
  Mechanical Engineering\\
  University of Texas at Dallas\\
  Richardson, TX, 75080 \\
  \texttt{william.anderson@utdallas.edu} \\
}

\begin{document}

\date{}
\maketitle

\begin{abstract}

Turbulent flow physics regulates the aerodynamic properties of lifting surfaces, the thermodynamic efficiency of vapor power systems, and exchanges of natural and anthropogenic quantities between the atmosphere and ocean, to name just a few applications of contemporary importance. The space-time dynamics of turbulent flows are described via numerical integration of the non-linear Navier-Stokes equation -- a procedure known as computational fluid dynamics (CFD). At the dawn of scientific computing in the late 1950s, it would be many decades before terms such as ``carbon footprint'' or ``sustainability'' entered the lexicon, and longer still before these themes attained national priority throughout advanced economies. The environmental cost associated with CFD is seldom considered. Yet, large-scale scientific computing relies on intensive cooling realized via external power generation that is primarily accomplished through the combustion of fossil fuels, which leads to carbon emissions. This paper introduces a framework designed to calculate the carbon footprint of CFD and its contribution to carbon emission reduction strategies. We will distinguish between "hero" and "routine" calculations, noting that the carbon footprint of hero calculations -- which demand significant computing resources at top-tier data centers -- is largely determined by the energy source mix utilized. We will also review CFD of flows where turbulence effects are modeled, thus reducing the degrees of freedom. Estimates of the carbon footprint are presented for such fully- and partially-resolved simulations as functions of turbulence activity and calculation year, demonstrating a reduction in carbon emissions by two to five orders of magnitude at practical conditions. Beyond analyzing CO2 emissions, we quantify the benefits of applying CFD towards overall carbon emission reduction. The community's effort to avoid redundant calculations via turbulence databases merits particular attention, with estimates indicating that a single database could potentially reduce CO2 emissions by approximately O(1) million metric tons. Additionally, implementing CFD in the fluids industry has markedly decreased dependence on wind tunnel testing, which is anticipated to lead to CO2 emission reduction.
\end{abstract}


\section{Introduction}

The concept of the carbon footprint has become central to assessing environmental impacts in the natural environment literature \cite{wiedmann2008definition, protocol1997kyoto}. 
Accurate carbon footprint quantification helps society understand the environmental impact of human activities and underpins strategies and policies to combat climate change \cite{hertwich2009carbon}.
Analyses of carbon footprints are now extensively available in scholarly literature. 
For example, Usman and Radulescu \cite{usman2022examining}, along with others \cite{kelly2019water, arevalo2022mitigation}, showed that renewable energy sources, while not entirely devoid of greenhouse gas emissions, significantly reduce carbon footprints compared to traditional fossil fuels.
In urban development, the concept of ``urban carbon footprints'' focuses on greenhouse gas emissions from cities. 
Dodman (2009) highlighted that urban areas, although covering a small fraction of the Earth's surface, contribute substantially to the global carbon footprint \cite{dodman2009urban}. 
Kissinger et al. (2013) explored the material consumption aspects of urban carbon footprints \cite{kissinger2013accounting}. 
Further work can be found in Refs. \cite{lin2015tracking, fan2012embedded,koide2021exploring} but is not detailed here. 
In agriculture, Adom et al. (2012) reported greenhouse gas emissions from common dairy feeds in the United States (US) \cite{adom2012regional}.
In public health, Okeke (2022) examined the impact of human carbon footprints, with an emphasis on Africa \cite{okeke2022carbon}.
Furthermore, Kanemoto et al. \cite{kanemoto2016mapping}, Ivanova et al. \cite{ivanova2017mapping}, Yang et al. \cite{yang2020mapping}, and Lomas et al. \cite{lomas20104m} estimated the carbon footprint mapping for the US, EU, China, and UK, providing spatial representation of carbon emissions and carbon sinks, taking factors like land use, international trades, supply chain, urbanization, and consumption patterns, into account. 

This study examines the carbon footprints associated with computational simulations.
Computational simulations serve as a crucial cornerstone in modern scientific research, enabling the exploration of complex phenomena across diverse fields such as physics, engineering, biology, and environmental science. 
By mimicking the behaviors of systems under various conditions, simulations offer invaluable insights into processes that are otherwise too challenging to observe directly, significantly reducing the time, cost, and ethical issues associated with traditional experiments. 
Here, we focus on computational fluid dynamics (CFD). 
CFD concerns the simulation and analysis of fluid flow, heat transfer, and related phenomena. 
By solving the Navier-Stokes equations numerically, CFD enables the prediction of fluid behavior in complex systems without the need for physical prototypes, reducing both cost and development time.


Recent advancements in high-performance computing (HPC), including petascale and exascale systems have significantly enhanced the accessibility of CFD \cite{bronevetsky2012reliable,kothe2018exascale}. 
Tools like large-eddy simulation (LES) have gained widespread application in industries like wind energy \cite{tamura2008towards,mehta2014large,abkar2015influence}, notably in validating wind farm models \cite{starke2021area,shapiro2018wind} and studying the interaction of wind turbines with the atmosphere \cite{Abkar2016,starke2023dynamic, narasimhan2022effects}. 
Additionally, LES has been instrumental in urban environment studies \cite{toja2018review} and in aerodynamics, particularly in the development of the ``certificate by analysis'' (CbA) concept \cite{mauery2021guide}, which seeks to replace part and even all of wind tunnel tests with CFD simulations. 
In this context, researchers such as Lehmkuhl et al. \cite{lehmkuhl2018large}, and others \cite{goc2023wind, goc2022large, goc2021large}, have successfully validated their CFD calculations against experimental drag and lift data. 
The application of CFD extends beyond these areas to include turbomachinery \cite{xu2022direct, corbett2024large, zhao2021high}, underwater hydrodynamics \cite{kroll2022large, morse2023tripping, plasseraud2023large}, and many other fields \cite{altland2022flow}.
Such widespread use of CFD leads to significant energy consumption and therefore a carbon footprint.
This paper estimates the carbon footprint of CFD and assesses CFD's role in mitigating carbon emissions.

The rest of the paper is organized as follows. 
The methodology is summarized in Sec. \ref{sec:method}.
The results and our discussion are presented in Sec. \ref{sec:DNS}.
Finally, we conclude in Sec. \ref{sec:conclusion}.

\section{Methods}
\label{sec:method}

The execution of CFD calculations relies on HPC centers, through which CFD has a carbon footprint. 
In this study, we focus on the carbon footprint during the execution of CFD only, and by adapting the method outlined in Ref. \cite{Lannelongue2023}, the carbon footprint of a CFD calculation is given by
\begin{equation}
\begin{split}
    {\rm Carbon \ footprint} = {\rm Energy \ consumption \times Carbon \ intensity}.
\end{split}
\label{eq:CF}
\end{equation}
Matthews et al. (2008) \cite{doi:10.1021/es703112w}, along with others \cite{tukker2000life,hauschild2018life,pennington2004life}, emphasized the importance of life-cycle analysis (LCA) for a holistic view encompassing the entire product life cycle. However, LCA's complexity warrants separate, future investigation. 

\begin{figure}
    \centering
    \includegraphics[width=0.4\textwidth]{ 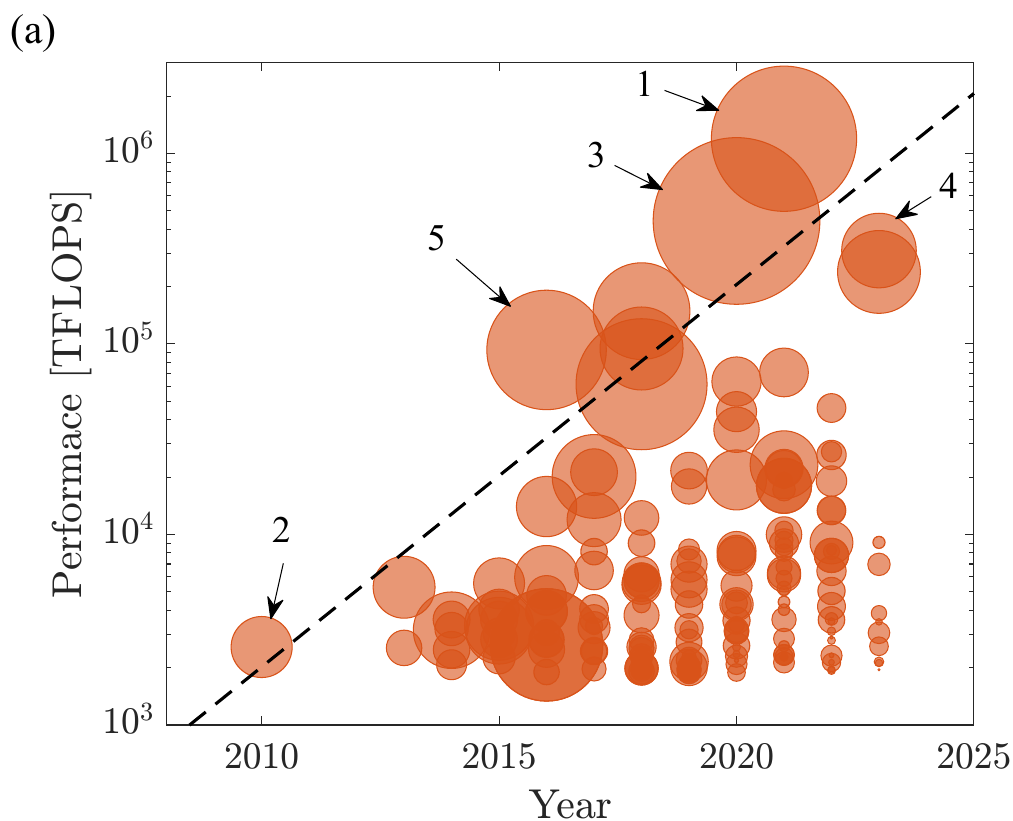}
    \includegraphics[width=0.4\textwidth]{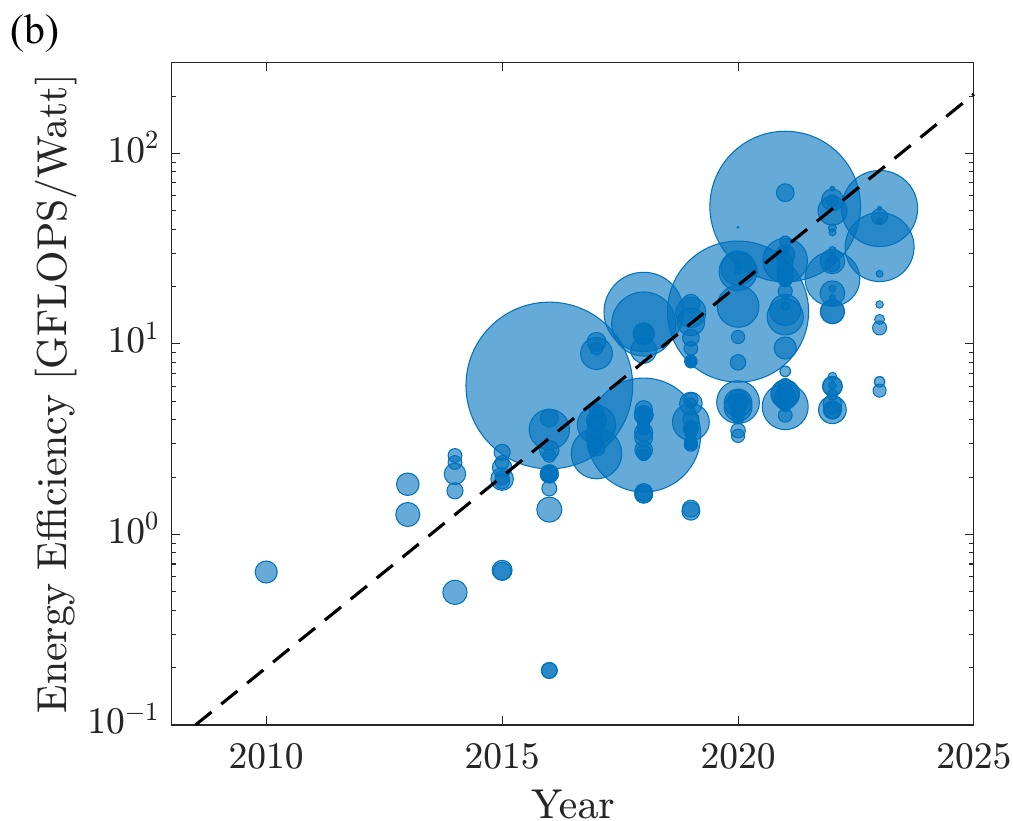}
    \caption{(a) Performance in FLOPS of the top 500 HPC systems \cite{Top500}.
    The circles represent the HPC systems. Their sizes represent the total power usage. 
    (b) Energy efficiency in GFLOPS/Watt of the top 500 HPC systems \cite{Top500}.
    The sizes of the circles represent the total core numbers of the centers.
    The dashed lines represent the doubling time of 1.5 years.
    The highlighted HPC systems are: 1. Frontier, 2. Tianhe-1A, 3. Supercomputer Fugaku, 4. Lumi, 5. Sunway Taihu-Light. }
    \label{fig:hpc}
\end{figure}

Equation \ref{eq:CF} contains two terms.
The energy consumption term is 
\begin{equation}
\begin{split}
    {\rm Energy \ Consumption}={\rm  Run~time  \times PSF \times IT \ power \ usage \times PUE}.
\end{split}
\label{eq:EC}
\end{equation}
Here, ``run time'' is the duration of a CFD calculation.
``PSF'' is the pragmatic scaling factor.
In the context here, PSF accounts for small-scale test runs.
For proficient CFD practitioners conducting large-scale CFD on high-end HPC clusters, the value of PSF would be very close to 1. 
``IT power usage'' is defined as
\begin{equation}
{\rm IT \ power \ usage} = \frac{\rm Performance}{\rm Energy \ efficiency}.
\label{eq:IT}
\end{equation}
``Performance'' is measured in floating-point operations per second (FLOPS), and energy efficiency in floating-point operations per Joule or FLOPS/Watt.
The peak performance and the energy efficiency of the top 500 HPC systems are displayed in Fig. \ref{fig:hpc}.
``Frontier'', which was put into service in 2021, is a top-tier HPC system.
It has a peak performance of $10^6$ TFLOPS and an energy efficiency of 52.59 GFLOPS/Watt.
Here, 1 TFLOPS is $10^{12}$ floating-point operations per second, and 1 GFLOPS is $10^9$ floating-point operations per second.
``Tianhe-1A'' is a low-tier HPC system (by today's standard).
Its performance is $2.5\times10^3$ TFLOPS, and its energy efficiency is 0.64 GFLOPS/Watt.
The ``Mira'' system, utilized for the $Re_\tau=5200$ channel flow DNS in 2013 \cite{lee2013petascale}, has a power usage of 3,945 kilowatts.
It was once a top contender but is now off the list, due to the rapid development of HPC. 
The data in Fig. \ref{fig:hpc} suggests that the performance and the energy efficiency of top-tier HPC systems have been doubling every one and a half years or so, following Moore's law.
The performance of the lower-tier systems, on the other hand, has stayed at 2$\times 10^3$ TFLOPS since 2010, with their energy efficiency increasing annually.

\begin{figure}
    \centering
    \includegraphics[width=0.4\textwidth]{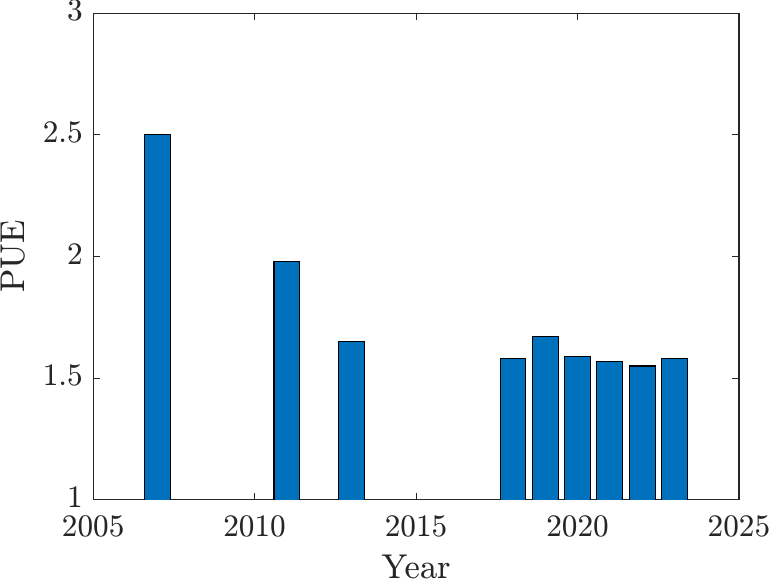}
    \caption{Annualized PUE from 2007 to 2023 \cite{uptimeinstitute}.}
    \label{fig:PUE}
\end{figure}

PUE or ``power usage effectiveness'' was a concept proposed by Malone and Belady \cite{malone2006metrics,GreenGrid}.
It measures the total energy needed to operate a data center and is defined as:
\begin{equation}
{\rm PUE}=\frac{\rm Facility \ Power \ Usage}{\rm IT \ Power \ Usage}.
\label{eq:PUE}
\end{equation}
The value of PUE depends on the IT equipment, the cooling system, the location, the climate, the size, and the design of the data center \cite{lei2022climate}.
A 2006 study of 22 US data centers reported PUE values between 1.33 and 3, averaging 2.04 \cite{greenberg2006best}.
A subsequent 2011 survey of 115 data centers, with 70 responses, indicated an average PUE of 1.69 \cite{kaiser2011survey}.
Figure \ref{fig:PUE} charts the evolution of the average PUE from 9 surveys conducted between 2006 and 2023 \cite{uptimeinstitute}.
The data suggests that the PUE value has stabilized at 1.5-1.7 since 2018.
Although more center-specific data is not available, the average value of PUE$=1.69$ suffices for the scope of this study.

\begin{figure}
    \centering
    \includegraphics[width=0.4\textwidth]{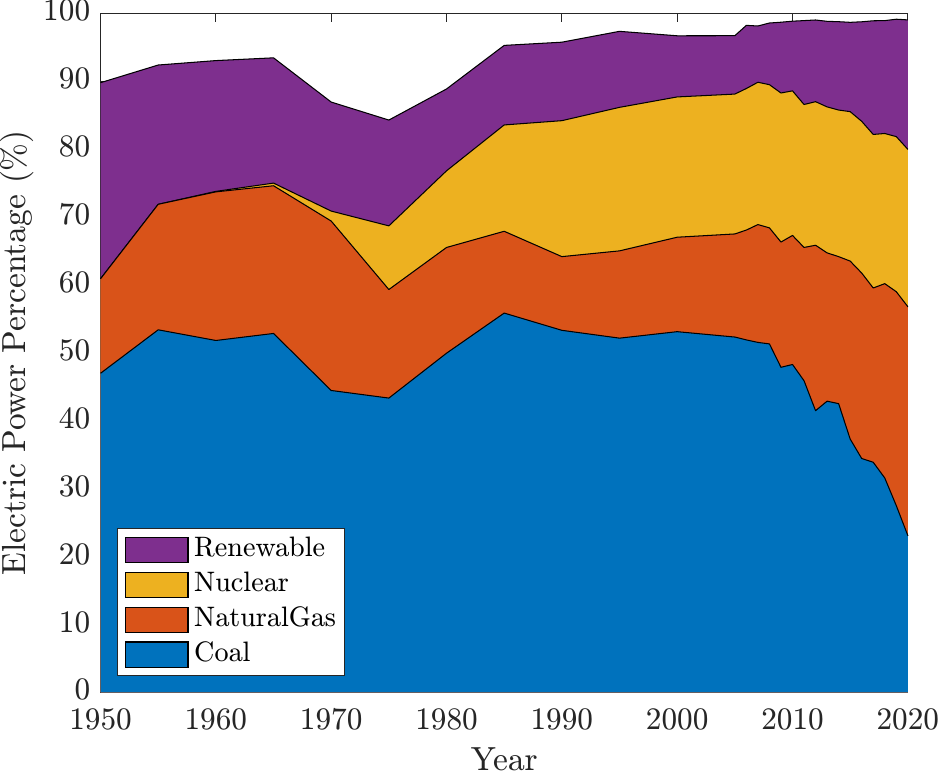}
    \caption{Composition of the energy sources in the US \cite{MonthlyEnergyReview}.}
    \label{fig:electricity}
\end{figure}

\begin{figure}
    \centering
    \includegraphics[width=0.4\textwidth]{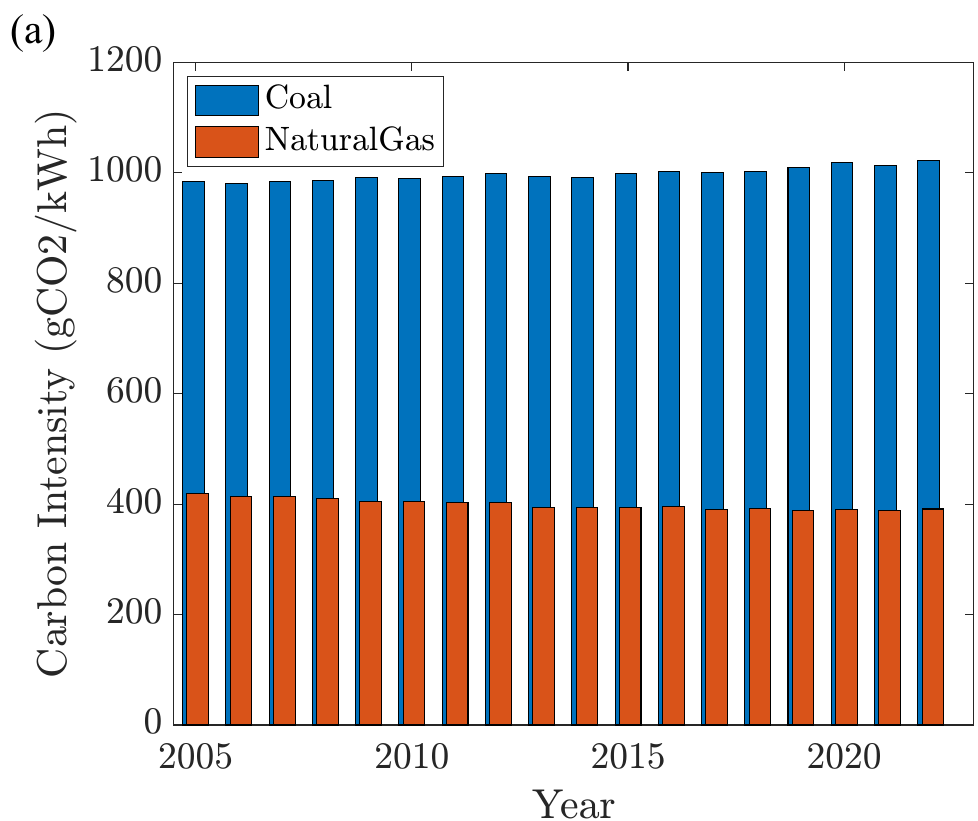}
    \includegraphics[width=0.4\textwidth]{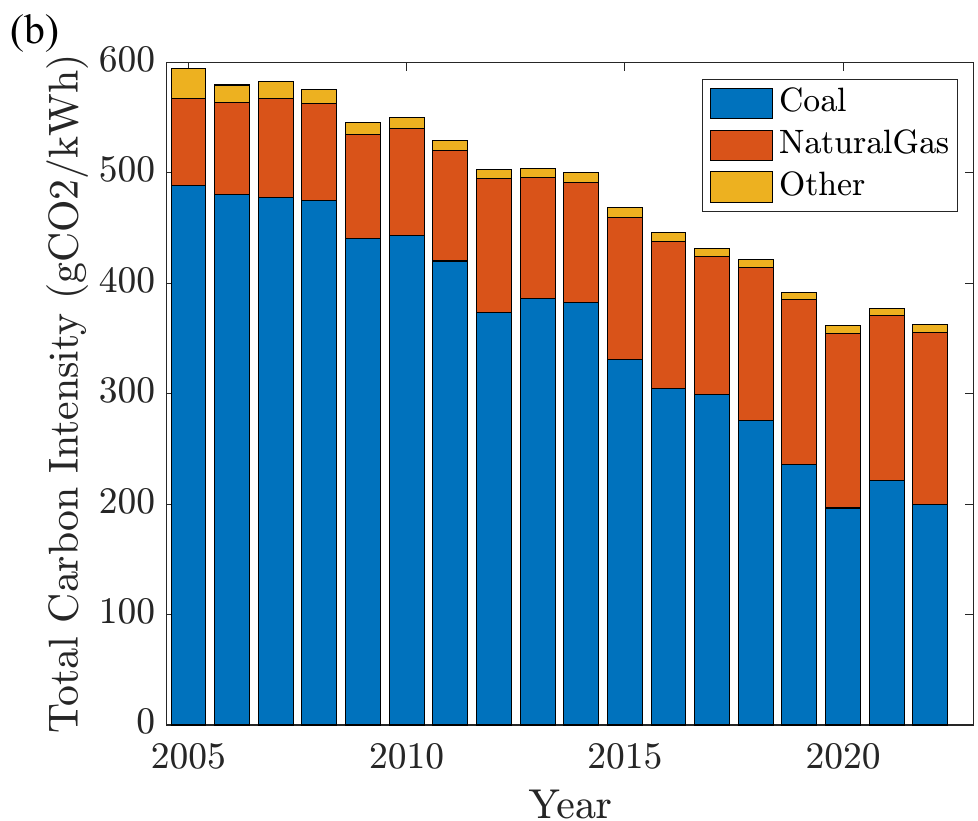}
    \caption{(a) Carbon intensity of coal and natural gas in the US \cite{MonthlyEnergyReview}.
    (b) Composition of the aggregated carbon intensity in the US \cite{ourworldindata}.}
    \label{fig:carbon_intensity}
\end{figure}

The carbon intensity (CI), quantifies the CO2 emissions per unit of energy produced. Expressed in grams of CO2 per kilowatt-hour (kWh), it is calculated as follows:
\begin{equation}
{\rm CI}=\sum_i p_i {\rm CI}_i.
\label{eq:CI}
\end{equation}
Here, CI represents the aggregated carbon intensity, $p_i$ denotes the fraction of the $i$th energy source in the overall mix, and CI$_i$ is the carbon intensity of that specific energy source. 
The composition of energy sources in the US from 1950 to 2022 is displayed in Fig. \ref{fig:electricity}.
Initially, coal, natural gas, and renewables were predominant until the 1980s, when nuclear power emerged as a significant contributor \cite{MonthlyEnergyReview}.
A notable shift is observed in the fraction of renewables, decreasing from 28.9\% in 1950 to 8.3\% in 2007, overshadowed by the rapid expansion of other energy forms. 
Additionally, coal's share in the energy mix plummeted from 55.9\% in 1985 to 23.1\% in 2020. 
The carbon intensities of the primary energy sources and the aggregated carbon intensity are displayed in Fig. \ref{fig:carbon_intensity}.
The carbon intensities of nuclear and renewable sources are relatively low, under 50 gCO2/kWh, and are negligible compared to those of coal and natural gas. 
The carbon intensity of coal has remained around 1000 gCO2/kWh, and natural gas at 400 gCO2/kWh. 
The overall carbon intensity has been on a steady decline, as indicated by the data. 
This reduction can be largely attributed to the decreasing reliance on coal and natural gas for electricity generation.

Combining Eqs. \ref{eq:CF}, \ref{eq:EC}, \ref{eq:IT}, and \ref{eq:CI}, we have the following estimate for the carbon footprint:
\begin{equation}
{\rm Carbon \ footprint} \approx {\rm Run \ time }\times \frac{\rm Performance}{\rm Energy \ efficiency}\times {\rm PUE}\times {\rm PSF} \times \left(p_{\rm coal} {\rm CI}_{\rm coal} + p_{\rm natural \ gas} {\rm CI}_{\rm natural \ gas}\right).
\label{eq:CF-gen}
\end{equation}
Here, we have neglected the emissions from nuclear and renewable energies. Based on the discussions in this section, PSF is  1, and PUE is  1.7. 
The carbon intensities of coal and natural gas have remained relatively stable since 2005 at 1000 gCO2/kWh and 400 gCO2/kWh, respectively.

\section{Results and discussion}
\label{sec:DNS}

\subsection{Hero and routine calculations}
We differentiate between ``hero'' and ``routine'' calculations. 
A hero calculation is done on a top-tier HPC system, utilizing a substantial portion of its cores for a stipulated period of time (stipulated by the data center). 
For these top-tier HPC systems, both the performance and the energy efficiency double approximately every 1.5 years. 
As a result, the performance-to-energy efficiency ratio remains approximately constant year-on-year, at around 10 MW. 
It then follows from Eq. \ref{eq:CF-gen} that the carbon footprint of a hero calculation is:
\begin{equation}
{\rm Carbon \ footprint}~({\rm kg~CO2}) \approx 
1.7\times 10^4 \times {\rm Run \ time }~(h) \times  \left(p_{\rm coal} + 0.4p_{\rm natural \ gas}\right),
\label{eq:CF-hero}
\end{equation}
where the run time usually does not exceed a couple of hundred hours, and $p_{\rm coal}$ and $p_{\rm natural gas}$ are the fraction of coal and natural gas in the mix of the energy sources.
This estimate tends to be on the higher side, assuming full cluster utilization. 
While not included here, an equivalent run time could adjust for this overestimation. 
Equation \ref{eq:CF-hero} implies that the carbon footprint of a hero calculation largely depends on the energy source mix. 
As the US shifts away from coal, the carbon footprint of such calculations is set to decrease.
Conversely, for a routine calculation on a low-tier HPC system, the run time can range from a few hours to a year, depending on the task and the availability of the resources. 
For low-tier, performance has stabilized at about $2\times10^3$ TFLOPS since 2010, and the energy efficiency doubles every 1.5 years or so. 
Thus,  Eq. \ref{eq:CF-gen} gives the following estimates for the carbon footprint of a routine CFD calculation:
\begin{equation}
{\rm Carbon \ footprint}~({\rm kg~CO2})\approx 6.6\times 10^3\times {\rm Run \ time }~({\rm h}) \times \left(p_{\rm coal}+0.4p_{\rm natural \ gas}\right)\times {2^{-\dfrac{{\rm Year}-2010}{1.5}}},
\label{eq:CF-routine}
\end{equation}
where we have baselined against Tianhe-1A established in 2010.
This too is an overestimate, assuming full cluster utilization. 
Aside from the energy source composition, the increase in energy efficiency will also contribute to the decrease of the carbon footprint.

\subsection{Direct numerical simulation}

Direct numerical simulation (DNS)  provides accurate solutions to the Navier-Stokes.
These solutions have yielded significant insights into flow physics, leading to the development of concepts such as the inner cycle \cite{Jimenez1999}, turbulence islands \cite{WuXiaohua2017}, and the momentum cascade \cite{Yang2017,chen2019non}.
Furthermore, DNS has been instrumental in validating models and theories, including the attached eddy model \cite{Marusic2019,yang2019hierarchical,yang2020scaling} and wall models \cite{yang2015integral,fowler2022lagrangian,vadrot2023survey}.
Consider, e.g., direct numerical simulation (DNS) of channel flow, a canonical configuration in fluid dynamics.
DNS of this flow dates back to 1987.
The calculation was at a friction Reynolds number of $Re_\tau=180$ in a domain sized $4\pi  \times 2 \times 2\pi $ \cite{Kim1987}. 
Subsequently, DNSs of channel flow at increasingly higher Reynolds numbers have been performed on progressively more powerful HPC systems. 
Table \ref{tab:DNS} summarizes some of the hero calculations, including their Reynolds numbers, computation times, HPC systems used, and domain and grid sizes.

\begin{table*}
\small
\centering
\begin{tabular}{ccccccccc}\hline
Year & $Re_{\tau}$ & Domain & Grid  & Cores & Wall time (h) & Region &Reference\\
1987 & 180 & $4\pi h \times 2h \times 2\pi h$ & $192\times129\times160$  & 4 & 62.5 & USA & Kim et al. \cite{Kim1987} \\
1999 & 590 & $2\pi h \times 2h \times \pi h$  & $384\times257\times384$  & N/A & N/A & USA & Moser et al. \cite{Moser1999}\\
2004 & 950 & $8\pi h \times 2h \times 3\pi h$ & $3072\times385\times2304$  & N/A & N/A & Spain/USA &  del Alamo et al. \cite{delAlamo2004}\\
2006  & 2k & $8\pi h \times 2h \times 3\pi h$ & $6144\times633\times4608$   & 2048 & $2929$ & Spain &  Hoyas \& Jimenez \cite{Hoyas2006} \\
2013& 5.2k & $8\pi h \times 2h \times 3\pi h$ & $10240\times1536\times7680$  & 524288 & 496 & USA &  Lee et al. \cite{lee2013petascale, LeeMoser2015}\\
2014  & 4k & $6\pi h \times 2h \times 2\pi h$ & $8192\times1024\times4096$ & N/A & N/A & Germany &  Bernardini et al. \cite{Bernardini2014}\\
2014& 4.2k & $2\pi h \times 2h \times \pi h$  & $2048\times1081\times2048$ & N/A & N/A & Spain &  Lozano \& Jimenez \cite{Lozano2014}\\
2018  & 8k & $16 h \times 2h \times 6.4 h$    & $8640\times4096\times6144$ & N/A & N/A & Japan &  Yamamoto \& Tsuji \cite{Yamamoto2018}\\
\hline
\end{tabular}
\caption{Details of some hero channel flow DNSs.
N/A is ``not available''.
}
\label{tab:DNS}
\end{table*}

\begin{figure*}
    \centering
    \includegraphics[width=1\textwidth]{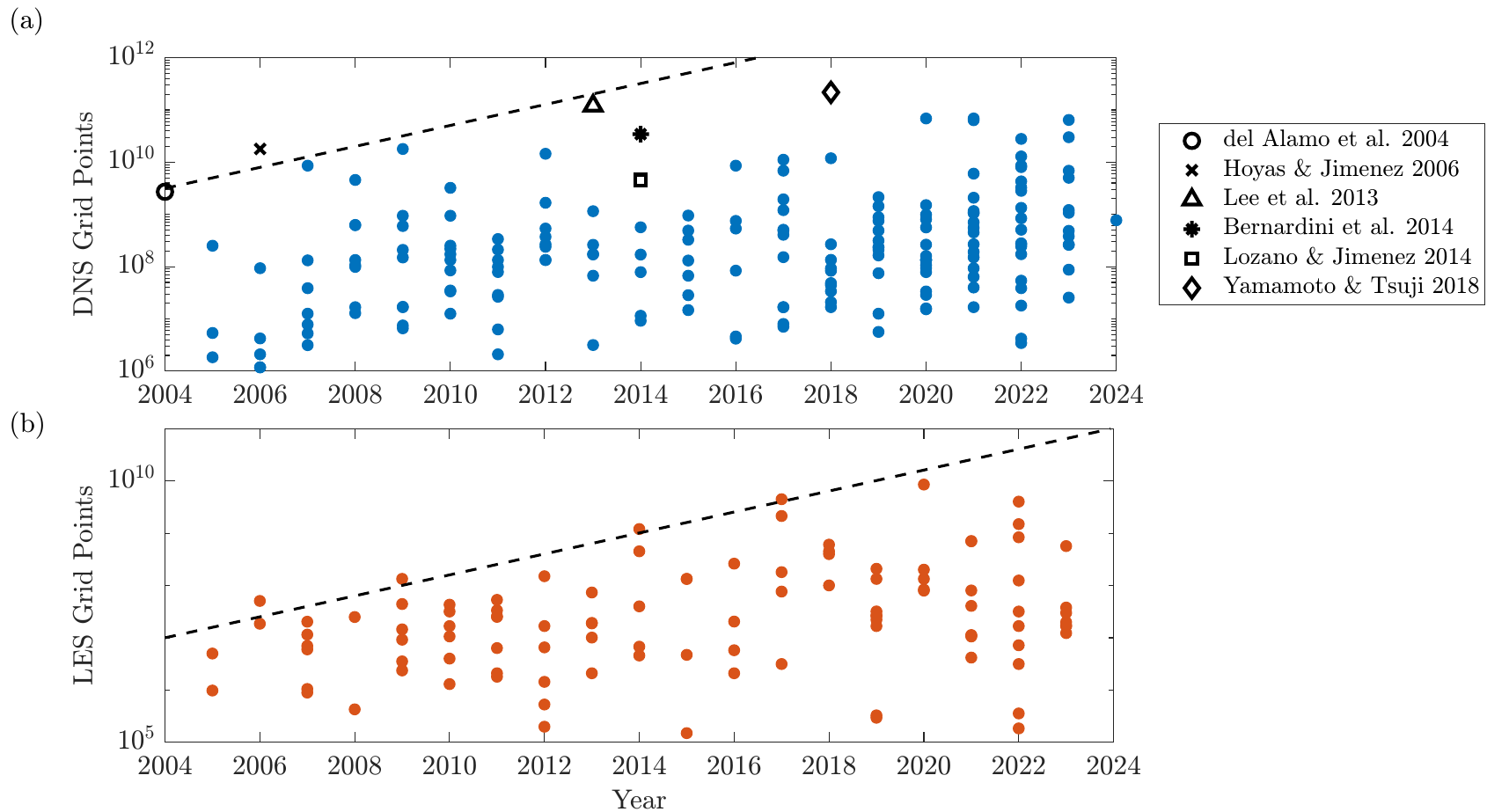}
    \caption{(a) The number of grid points in DNSs. (b) The number of grid points in LESs.
    The hero calculations in Table \ref{tab:DNS} are highlighted.
    The dots are DNS and LES reported in the Journal of Fluid Mechanics. 
    The references used to generate the figure are detailed in the Appendix.
    This is not meant to be comprehensive.
    The dashed lines represent a doubling time of 1.5 years.}
    \label{fig:channel}
\end{figure*}

Figure \ref{fig:channel} illustrates the sizes of the DNS calculations in terms of grid points.
Here, we compile data for DNSs as well as LES published in the Journal of Fluid Mechanics.
The figure contains the hero calculation data in Table \ref{tab:DNS}, and routine calculations as well.
The size of the hero calculations has escalated from fewer than $10^7$ grid points in 1987 \cite{Kim1987} to over $10^{11}$ in 2015 \cite{lee2013petascale}, yet there has not been a significant increase since.
The sizes of routine calculations, on the other hand, range from $10^6$ to the size of the hero calculation of the time for DNS and from $10^5$ to the size of the hero calculation of the time for LES.
In these two figures, we have indicated the expected growth if the CFD community had kept pace with HPC developments, i.e., running the same DNS and LES codes on the top machines.
The trend, shown in Fig. \ref{fig:channel}, is somewhat disheartening. 
It appears the CFD community has not continued to invest in hero calculations at the rate of HPC developments.

We now present a case study and estimate the carbon footprint of the channel DNS in Ref. \cite{lee2013petascale}. 
This simulation was performed on the ``Mira'' HPC system.
The calculation utilized 524k cores out of the 768k cores in the system and lasted for 500 hours--- it is one of the largest CFD calculations to date. 
In the year 2013, the fraction of coal and natural gas in the mix of the energy sources are 42.9\% and 21.8\%.
It follows from Eq. \ref{eq:CF-gen} that the carbon footprint of this DNS calculation is:
\begin{equation}
{\rm Carbon \ footprint} \approx 500 {\rm ~h} \times 4 {\rm~MW}\times \frac{524}{768} \times 1.69
\times 1 \times (42.9\% + 0.4\times21.8\%) \text{~kg CO2/kWh} \approx 10^6 {\rm ~kgCO2}.
\end{equation}
To contextualize this figure, the carbon footprint of a Boeing 777 flight from New York to Beijing and back is approximately $1.4 \times 10^6$ kgCO2, assuming 3600 kg CO2 emission per passenger and 388 passengers on the plane \cite{CO2cal}.

Before we proceed, we make a remark.
It is common that a paper elaborates on the numerics and the setup but says nothing about the actual computation, including the computing time and the cluster used. 
This shows a lack of awareness of the issue of carbon footprint.
A goal of this work is to help bring this aspect of CFD to the reader's attention.

\begin{figure*}
    \centering
    \includegraphics[width=0.8\textwidth]{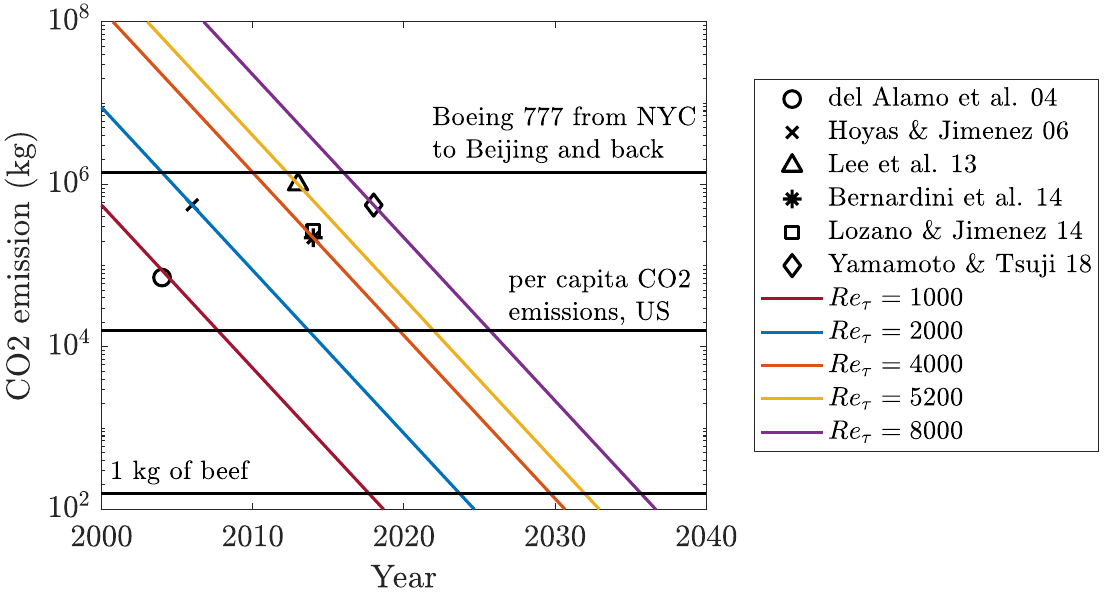}
    \caption{Estimates of carbon footprint for channel flow DNS according to Eq. \ref{eq:cf-chan-e}.}
    \label{fig:CF-chan}
\end{figure*}

Although it is hard to get direct estimates for carbon emissions due to a lack of information, indirect estimates are viable. 
Here, we provide such estimates as a function of the Reynolds number and the year of calculation. 
We will assume constant carbon intensity and resource utilization efficiency.
Errors due to these assumptions are not to exceed a factor of 2---considering the rapid annual increase of energy efficiency and FLOP counts, such errors are acceptable.
Utilizing Eq. \ref{eq:CF-gen}, we have
\begin{equation}
\label{eq:CF-chan}
{\rm Carbon \ footprint} \approx {\rm Carbon \ footprint}_{\rm baseline}\times\frac{{\rm FLOP \ counts}}{\left.{\rm FLOP \ counts}\right|_{\rm baseline}}
\frac{\left.{\rm Energy \ efficiency}\right|_{\rm baseline}}{{\rm Energy \ efficiency}}.
\end{equation}
That is, the carbon footprint of a calculation is proportional to the FLOP counts and inversely proportional to the energy efficiency.
The FLOP count scales as $Re_\tau^4$ for DNS of channel flow, assuming fixed domain size, grid resolution, the Courant–Friedrichs–Lewy number \cite{yang2021grid}:
\begin{equation}
\frac{{\rm FLOP \ counts}}{\left.{\rm FLOP \ counts}\right|_{\rm baseline}} = \left(\frac{Re_\tau}{Re_{\tau,{\rm baseline}}}\right)^4.
\label{eq:FLOP-r}
\end{equation}
The energy efficiency ratio is
\begin{equation}
\frac{\left.{\rm Energy \ efficiency}\right|_{\rm baseline}}{{\rm Energy \ efficiency}}
={2^{-\dfrac{{\rm Year}-{\rm Baseline \ Year}}{1.5}}},
\label{eq:EE-r}
\end{equation}
per Moore's law.
By taking the $Re_\tau=5200$ channel flow calculation as our baseline, the above equations yield
\begin{equation}
{\rm Carbon \ footprint} ({\rm kg~CO2}) \approx 10^6\times \left(\frac{Re_\tau}{5200}\right)^4\times{{2}^{-\dfrac{{\rm Year}-2013}{1.5}}}.
\label{eq:cf-chan-e}
\end{equation}
Figure \ref{fig:CF-chan} illustrates the carbon footprint of channel flow DNS calculations at varying $Re_\tau$ values, from 1000 to 8000, between year 2000 to year 2040.
To provide context, we indicate the carbon emission of serving 1 kg of beef \cite{poore2018reducing}, the per capita CO2 emission in the US in the year 2022 \cite{statistaCO2Emissions2023}, as well as the emissions from a Boeing 777 flight from New York to Beijing and back \cite{CO2cal}.
From Fig. \ref{fig:CF-chan}, we see that the carbon footprint escalates significantly with the Reynolds number. 
For instance, a $Re_\tau=1000$ channel flow calculation today emits less CO2 than producing 1 kg of beef, whereas a $Re_\tau=8000$ calculation's emissions are on par with those of a round-trip Boeing 777 flight from New York to Beijing.
Nonetheless, for a given Reynolds number, the carbon footprint decreases annually, thanks to the increasing energy efficiency. 
For example, repeating the $Re_\tau=5200$ channel flow DNS in 2025 would result in a carbon footprint of $10^4$ kg CO2. 
Lastly, examining ``hero'' calculations from 2000 to 2020, we observe a plateau and a slight decrease in the carbon footprint.
This is consistent with the results in Fig. \ref{fig:channel}---the CFD community has not kept up with the developments of HPC.

\subsection{Turbulence modeling}
DNS is prohibitively expensive for flows at practically relevant Reynolds numbers due to the necessity of resolving all scales within the flow. To make CFD practical, one approach is to model the small scales and the flow within the wall layer. This gives rise to wall-resolved LES (WRLES) and wall-modeled LES (WMLES).

In this subsection, we explore the benefits of turbulence modeling through the lens of carbon emissions reduction. While current literature on LES lacks the data to directly estimate the carbon footprint, indirect estimation methods are feasible. We employ Equation \ref{eq:CF-chan} for this purpose. The energy efficiency ratio is determined by Eq. \ref{eq:EE-r}, and we assume that the ratio of floating-point operation (FLOP) counts mirrors the computational cost ratio. To further simplify, we assume the computational cost directly correlates with grid size. This implies equivalent time steps for LES and DNS, which is a crude approximation.

According to analyses in Refs. \cite{yang2021grid, choi2012grid}, we derive the following FLOP count ratios:
\begin{equation}
\frac{{Re_{L_x}=3.5e7 {\rm ~BL~DNS}}}{{Re_\tau=5200~{\rm channel~DNS}}} \approx 1,
\label{eq:FC-r1}
\end{equation}
and
\begin{equation}
\begin{split}
\frac{{Re_{L_x}~{\rm BL~DNS}}}{{Re_{L_x}=3.5e7 {\rm ~BL~DNS}}} &\approx \left(\frac{Re_{L_x}}{3.5e7}\right)^{2.05},\\
\frac{{Re_{L_x}~{\rm BL~WRLES}}}{{~Re_{L_x}=3.5e7 {\rm ~BL~DNS}}} &\approx 0.68\frac{Re_{L_x}^{1.86}}{(3.5e7)^{2.05}},\\
\frac{{Re_{L_x}~{\rm BL~WRLES}}}{{Re_{L_x}=3.5e7 {\rm ~BL~DNS}}} &\approx 7.6e4\frac{Re_{L_x}^{1.00}}{(3.5e7)^{2.05}}.\end{split}
\label{eq:FC-r2}
\end{equation}
These ratios, and hence the subsequent analysis, rest on the assumptions delineated in Ref. \cite{choi2012grid}. From Eqs. \ref{eq:CF-chan}, \ref{eq:EE-r}, \ref{eq:FC-r1}, and \ref{eq:FC-r2}, we infer the carbon footprints for boundary layer DNS, WRLES, and WMLES as follows:
\begin{equation}
\text{Carbon footprint of boundary layer DNS}~({\rm kg~CO2})\approx  10^6 \times \left(\frac{Re_{L_x}}{3.5e7}\right)^{2.05}\times {{2}^{-\dfrac{{\rm Year}-2013}{1.5}}},
\label{eq:cf-bl-dns}
\end{equation}
\begin{equation}
\text{Carbon footprint of boundary layer WRLES}~({\rm kg~CO2}) \approx 7\times 10^5\times \frac{Re_{L_x}^{1.86}}{(3.5e7)^{2.05}}\times {{2}^{-\dfrac{{\rm Year}-2013}{1.5}}},
\label{eq:cf-bl-wrles}
\end{equation}
and
\begin{equation}
\text{Carbon footprint of boundary layer WMLES}~({\rm kg~CO2}) \approx 10^{11}\times \frac{Re_{L_x}^{1}}{(3.5e7)^{2.05}}\times {{2}^{-\dfrac{{\rm Year}-2013}{1.5}}}.
\label{eq:cf-bl-wmles}
\end{equation}
Here, we have assumed identical efficiency for LES and DNS codes and omitted the costs related to sub-grid scale and wall models. 
These assumptions likely make our estimates conservative. Further refinement of these estimates is challenging and is deferred to future research.
For boundary layers at a Reynolds number of $Re_{L_x}=10^9$, typical for underwater vehicles, our calculations yield carbon footprint estimates for DNS, WRLES, and WMLES in 2023 of $10^7$, $10^5$, and $3\times10^2$ kg, respectively. These figures starkly illustrate that DNS at such Reynolds numbers is impractical, WRLES, which models the sub-grid scale motions, is feasible but with significant carbon emissions, and WMLES, which models both the sub-grid scale motions and the near-wall flows, offers a viable and environmentally sustainable option.

\subsection{Turbulence Database}
High-fidelity CFD simulations, known for their computational intensity, necessitate significant resources and result in notable carbon emissions. 
To mitigate these demands, various online repositories have been established, hosting extensive datasets from simulations. 
Here, we take the Johns Hopkins Turbulence Database (JHTDB) as an example \cite{li2008public, graham2016web}.
The database offers DNS data on isotropic turbulence, magneto-hydrodynamic turbulence, channel flow, and boundary layer flows, encompassing over 300 Terabytes of data \cite{JHTDB}.
To date, the database has received $4\times 10^{14}$ point queries.
The carbon footprint of the CFD solution at a grid point is not straightforward to accurately estimate, as it involves flow condition, the cluster of the calculation, and the time of the calculation, among others.
Nonetheless, an approximate figure can be drawn from the analyses of the $Re_\tau=5200$ channel flow DNS, i.e., our baseline case.
Per the estimate there, generating CFD solution at one grid point leads to a carbon emission of $8.3\times 10^{-3}$ g CO2.
Considering that the power needed to operate the database is negligible compared to that needed for CFD calculation, JHTDB alone has contributed to a $3.3$ million metric tons CO2 reduction in carbon emission.


\subsection{Aviation}

In the field of aviation, CFD has emerged as a standard tool alongside wind tunnel. 
Its use in the process of design, analysis, certification, and maintenance has facilitated progress in both internal and external flows \cite{spalart2016role,mani2023perspective} reducing the reliance on physical wind tunnel testing and at the same time accelerating the development cycles of aircraft. 
Johnson et al. (2005) highlighted a significant reduction in the number of wind tunnel tests for aircraft wings at Boeing, from 77 in the late 1970s to just 10 by the late 1990s \cite{johnson2005thirty}, i.e., a reduction of 87\%.
This shift diminished the need for extensive wind tunnel hours, through which CFD helps reduce carbon emissions \cite{poisson1968wind}.
Take the Boeing 747 as an example.
Its development in the 60s and the 70s relied primarily on wind tunnel experiments.
The wind tunnel hours were about 10,000 \cite{desai2003relative}. 
The power of the fan in the Beoing low-speed aero acoustic facility is 7 MW \cite{boeingWindTunnels2024}.
It follows that the energy consumption of wind tunnel tests is $O(10^8)$ kWh.
It is not fair to compute energy saving according to 87\%$\times O(10^8)$ kWh, and it is difficult to estimate the reduction in carbon emission, but it is safe to claim the potential of CFD in reducing the carbon footprint of aircraft design.

CFD is also set to play a critical role in mitigating the aviation industry's carbon footprint. 
Without advancements in technology, CO2 emissions from domestic and international aviation are projected to reach 450 million metric tons by 2050 \cite{US2021action}.
The goal, as outlined in the US Aviation Climate Action Plan, is to stabilize carbon emissions at the 2019 level of approximately 210 million metric tons \cite{US2021action}, thus creating a reduction target of 240 million metric tons. 
Biofuels derived from saturated fatty acids are expected to address about half of this gap \cite{mofijur2023screening, pasa2022main}.
Operational improvements contribute a very small portion. 
The renewal of air fleets with newer, more efficient aircraft is anticipated to account for the remainder of the reduction, a process that is significantly influenced by the application of CFD in aircraft design and development.


\section{Conclusions}
\label{sec:conclusion}

We propose a framework for estimating the carbon footprint of computational simulations, incorporating factors such as runtime, performance, energy efficiency, power usage effectiveness (PUE), scaling efficacy, energy mix, and the carbon intensity of each energy source.
We delineate hero from routine calculations, noting that the carbon footprint of the former is directly linked to carbon intensity. As the United States moves away from coal, a reduction in the environmental impact of these calculations is anticipated. Conversely, the carbon footprint of routine calculations is influenced by both carbon intensity and energy efficiency, which are improving rapidly.

A case study on the $Re_\tau=5200$ channel flow DNS reveals an estimated carbon footprint of $10^6$ kg CO2, equivalent to emissions from a round-trip Boeing 777 flight between New York and Beijing. 
Based on this data, we derive formulas for estimating the carbon footprints of different types of DNS and LES, showing significant benefits of turbulence modeling.
These formulas will allow CFD practitioners to estimate the environmental impacts of their calculations as a function of the Reynolds number and the year--before or after the calculation.

The paper also highlights the CFD community's efforts in carbon emission reduction, such as the Johns Hopkins Turbulence Database (JHTDB), which has saved approximately 3.3 million metric tons of carbon emissions by eliminating redundant calculations. Moreover, CFD's role in minimizing wind tunnel tests has notably reduced carbon emissions.

Last but not least, we call for a balanced view.
Take CFD as an example, while it generates carbon emissions, CFD has also contributed to the reduction of carbon emissions---both sides must be considered when assessing the environmental impact.

\section*{Acknowledgement}

This work was presented at the Flow, Turbulence, and Wind Energy Symposium.
The authors acknowledge inputs from the participants.

\appendix
\section{Supporting material for Fig. 5}

References used to generate Fig. \ref{fig:channel}.

\begin{table}\centering
\small
\caption{References used for the data points in Figure 1 (Part 1/3)}
\begin{tabular}{llll|llll}
Year & Max Grid Points & DOI & Type & Year & Max Grid Points & DOI & Type \\
\midrule
2024 & 774400000 & 10.1017/jfm.2023.1099& DNS &2020 & 805306368 & 10.1017/jfm.2019.1012& DNS \\
2023 & 1073741824 & 10.1017/jfm.2023.971& DNS &2020 & 160168320 & 10.1017/jfm.2020.601& DNS \\
2023 & 6865551360 & 10.1017/jfm.2023.104& DNS &2020 & 33554432 & 10.1017/jfm.2020.402& DNS \\
2023 & 5033164800 & 10.1017/jfm.2023.26& DNS &2020 & 563871744 & 10.1017/jfm.2020.173& DNS \\
2023 & 30064771072 & 10.1017/jfm.2023.764& DNS &2020 & 78643200 & 10.1017/jfm.2020.33& DNS \\
2023 & 25600000 & 10.1017/jfm.2023.884& DNS &2020 & 28794880 & 10.1017/jfm.2020.452& DNS \\
2023 & 382205952 & 10.1017/jfm.2023.728& DNS &2020 & 989148160 & 10.1017/jfm.2020.669& DNS \\
2023 & 485587656 & 10.1017/jfm.2022.1056& DNS &2020 & 94371840 & 10.1017/jfm.2020.874& DNS \\
2023 & 64424509440 & 10.1017/jfm.2022.1013& DNS &2020 & 68719476736 & 10.1017/jfm.2020.146& DNS \\
2023 & 87512161 & 10.1017/jfm.2023.502& DNS &2020 & 15271410 & 10.1017/jfm.2020.629& DNS \\
2023 & 1212678144 & 10.1017/jfm.2023.359& DNS &2020 & 15884544 & 10.1017/jfm.2020.590& DNS \\
2023 & 263831040 & 10.1017/jfm.2023.870& DNS &2020 & 1509949440 & 10.1017/jfm.2020.488& DNS \\
2023 & 262144000 & 10.1017/jfm.2023.307& DNS &2020 & 263831040 & 10.1017/jfm.2020.412& DNS \\
2022 & 843750000 & 10.1017/jfm.2022.764& DNS &2020 & 134217728 & 10.1017/jfm.2020.218& DNS \\
2022 & 1337720832 & 10.1017/jfm.2022.559& DNS &2020 & 134217728 & 10.1017/jfm.2020.162& DNS \\
2022 & 54000000 & 10.1017/jfm.2022.22& DNS &2020 & 106168320 & 10.1017/jfm.2020.780& DNS \\
2022 & 8589934592 & 10.1017/jfm.2022.434& DNS &2019 & 5640192 & 10.1017/jfm.2019.482& DNS \\
2022 & 18000000 & 10.1017/jfm.2022.822& DNS &2019 & 1440000000 & 10.1017/jfm.2019.670& DNS \\
2022 & 243855360 & 10.1017/jfm.2022.749& DNS &2019 & 491520000 & 10.1017/jfm.2019.376& DNS \\
2022 & 268435456 & 10.1017/jfm.2022.402& DNS &2019 & 2147483648 & 10.1017/jfm.2018.1005& DNS \\
2022 & 12884901888 & 10.1017/jfm.2022.942& DNS &2019 & 210134016 & 10.1017/jfm.2019.787& DNS \\
2022 & 509607936 & 10.1017/jfm.2021.1080& DNS &2019 & 235376180 & 10.1017/jfm.2018.953& DNS \\
2022 & 3446016 & 10.1017/jfm.2022.587& DNS &2019 & 235929600 & 10.1017/jfm.2019.84& DNS \\
2022 & 3263299584 & 10.1017/jfm.2022.393& DNS &2019 & 75000000 & 10.1017/jfm.2019.179& DNS \\
2022 & 27917287424 & 10.1017/jfm.2022.574& DNS &2019 & 754974720 & 10.1017/jfm.2019.558& DNS \\
2022 & 280350720 & 10.1017/jfm.2022.456& DNS &2019 & 164167200 & 10.1017/jfm.2018.1000& DNS \\
2022 & 2828800000 & 10.1017/jfm.2022.80& DNS &2019 & 314572800 & 10.1017/jfm.2019.222& DNS \\
2022 & 4294967296 & 10.1017/jfm.2022.294& DNS &2019 & 884736000 & 10.1017/jfm.2019.995& DNS \\
2022 & 4194304 & 10.1017/jfm.2022.801& DNS &2019 & 12582912 & 10.1017/jfm.2019.100& DNS \\
2022 & 173232000 & 10.1017/jfm.2022.331& DNS &2018 & 33816576 & 10.1017/jfm.2018.242& DNS \\
2022 & 8053063680 & 10.1017/jfm.2022.699& DNS &2018 & 135075000 & 10.1017/jfm.2018.489& DNS \\
2022 & 38755584 & 10.1017/jfm.2021.1028& DNS &2018 & 11894784000 & 10.1017/jfm.2018.625& DNS \\
2021 & 707481600 & 10.1017/jfm.2021.827& DNS &2018 & 20966400 & 10.1017/jfm.2018.466& DNS \\
2021 & 150994944 & 10.1017/jfm.2021.814& DNS &2018 & 44160000 & 10.1017/jfm.2018.256& DNS \\
2021 & 93219840 & 10.1017/jfm.2021.524& DNS &2018 & 268435456 & 10.1017/jfm.2018.231& DNS \\
2021 & 1073741824 & 10.1017/jfm.2021.32& DNS &2018 & 83420000 & 10.1017/jfm.2018.503& DNS \\
2021 & 452984832 & 10.1017/jfm.2021.519& DNS &2018 & 92798976 & 10.1017/jfm.2018.408& DNS \\
2021 & 268435456 & 10.1017/jfm.2021.511& DNS &2018 & 16777216 & 10.1017/jfm.2018.389& DNS \\
2021 & 63606620160 & 10.1017/jfm.2021.231& DNS &2018 & 48500000 & 10.1017/jfm.2018.331& DNS \\
2021 & 39911040 & 10.1017/jfm.2021.448& DNS &2017 & 16777216 & 10.1017/jfm.2016.859& DNS \\
2021 & 1061683200 & 10.1017/jfm.2021.103& DNS &2017 & 1207959552 & 10.1017/jfm.2017.237& DNS \\
2021 & 16777216 & 10.1017/jfm.2021.757& DNS &2017 & 1944000000 & 10.1017/jfm.2017.53& DNS \\
2021 & 63700992 & 10.1017/jfm.2021.705& DNS &2017 & 152174592 & 10.1017/jfm.2017.372& DNS \\
2021 & 68719476736 & 10.1017/jfm.2021.288& DNS &2017 & 509607936 & 10.1017/jfm.2017.873& DNS \\
2021 & 554000000 & 10.1017/jfm.2021.312& DNS &2017 & 489715200 & 10.1017/jfm.2017.619& DNS \\
2021 & 192000000 & 10.1017/jfm.2021.535& DNS &2017 & 16777216 & 10.1017/jfm.2017.672& DNS \\
2021 & 2087321600 & 10.1017/jfm.2020.1144& DNS &2017 & 8000000 & 10.1017/jfm.2017.157& DNS \\
2021 & 1160000000 & 10.1017/jfm.2020.943& DNS &2017 & 408158208 & 10.1017/jfm.2017.398& DNS \\
2021 & 6039797760 & 10.1017/jfm.2020.913& DNS &2017 & 7077888 & 10.1017/jfm.2017.164& DNS \\
2020 & 134217728 & 10.1017/jfm.2020.159& DNS &2017 & 6879707136 & 10.1017/jfm.2017.371& DNS \\
\bottomrule
\end{tabular}
\end{table}

\begin{table}\centering
\small
\caption{References used for the data points in Figure 1 (Part 2/3)}
\begin{tabular}{llll|llll}
Year & Max Grid points & DOI & Type & Year & Max Grid points & DOI & Type \\
\midrule
2017 & 11158866000 & 10.1017/jfm.2017.549& DNS &2010 & 12582912 & 10.1017/S0022112010000169& DNS \\
2016 & 8589934592 & 10.1017/jfm.2015.754& DNS &2010 & 650000 & 10.1017/S002211201000340X& DNS \\
2016 & 4186161 & 10.1017/jfm.2016.346& DNS &2010 & 35000000 & 10.1017/S002211200999423X& DNS \\
2016 & 4608000 & 10.1017/jfm.2016.554& DNS &2010 & 3227516928 & 10.1017/S0022112010003113& DNS \\
2016 & 536870912 & 10.1017/jfm.2016.207& DNS &2009 & 17921212416 & 10.1017/S0022112009007769& DNS \\
2016 & 752640000 & 10.1017/jfm.2016.179& DNS &2009 & 943936000 & 10.1017/S0022112009991388& DNS \\
2016 & 83886080 & 10.1017/jfm.2016.230& DNS &2009 & 149921280 & 10.1017/S0022112009992333& DNS \\
2015 & 131072000 & 10.1017/jfm.2015.9& DNS &2009 & 16777216 & 10.1017/S0022112008004916& DNS \\
2015 & 28459008 & 10.1017/jfm.2015.717& DNS &2009 & 7490265 & 10.1017/S0022112009007496& DNS \\
2015 & 120795955200 & 10.1017/jfm.2015.268& DNS &2009 & 17040384 & 10.1017/S0022112008004813& DNS \\
2015 & 14745600 & 10.1017/jfm.2015.566& DNS &2009 & 209715200 & 10.1017/S0022112009006624& DNS \\
2015 & 491206500 & 10.1017/jfm.2014.678& DNS &2009 & 6600000 & 10.1017/S0022112008005156& DNS \\
2015 & 327680000 & 10.1017/jfm.2015.696& DNS &2009 & 600000000 & 10.1017/S0022112008004473& DNS \\
2015 & 952247081 & 10.1017/jfm.2014.715& DNS &2008 & 4574147680 & 10.1017/S0022112008000864& DNS \\
2015 & 67108864 & 10.1017/jfm.2015.211& DNS &2008 & 622000000 & 10.1017/S0022112008001006& DNS \\
2014 & 11500000 & 10.1017/jfm.2014.589& DNS &2008 & 532480 & 10.1017/S0022112008004060& DNS \\
2014 & 9200000 & 10.1017/jfm.2014.368& DNS &2008 & 134217728 & 10.1017/S0022112007009883& DNS \\
2014 & 170393600 & 10.1017/jfm.2014.597& DNS &2008 & 106000000 & 10.1017/S0022112007009561& DNS \\
2014 & 78643200 & 10.1017/jfm.2014.68& DNS &2008 & 12880000 & 10.1017/S0022112008001985& DNS \\
2014 & 567730944 & 10.1017/jfm.2014.30& DNS &2008 & 98991585 & 10.1017/S0022112008000657& DNS \\
2013 & 262144000 & 10.1017/jfm.2012.596& DNS &2008 & 16777216 & 10.1017/S0022112008000141& DNS \\
2013 & 171884160 & 10.1017/jfm.2013.70& DNS &2008 & 629145600 & 10.1017/S0022112008002085& DNS \\
2013 & 3162112 & 10.1017/jfm.2013.361& DNS &2007 & 8589934592 & 10.1017/S0022112007008002& DNS \\
2013 & 67108864 & 10.1017/jfm.2013.130& DNS &2007 & 131995521 & 10.1017/S0022112007006192& DNS \\
2013 & 1153433600 & 10.1017/jfm.2013.142& DNS &2007 & 7807488 & 10.1017/S0022112007007380& DNS \\
2013 & 172800000 & 10.1017/jfm.2013.108& DNS &2007 & 12582912 & 10.1017/S0022112007004971& DNS \\
2012 & 268435456 & 10.1017/jfm.2012.59& DNS &2007 & 5270000 & 10.1017/S0022112007005836& DNS \\
2012 & 240000000 & 10.1017/jfm.2012.345& DNS &2007 & 3145728 & 10.1017/S0022112006004034& DNS \\
2012 & 134217728 & 10.1017/jfm.2012.241& DNS &2007 & 38707200 & 10.1017/S0022112007009020& DNS \\
2012 & 14495514624 & 10.1017/jfm.2012.428& DNS &2006 & 4227072 & 10.1017/S0022112006001121& DNS \\
2012 & 134217728 & 10.1017/jfm.2012.374& DNS &2006 & 2097152 & 10.1017/S0022112006001832& DNS \\
2012 & 373800960 & 10.1017/jfm.2012.400& DNS &2006 & 1179648 & 10.1017/S0022112006000711& DNS \\
2012 & 536870912 & 10.1017/jfm.2012.81& DNS &2006 & 93400000 & 10.1017/S002211200600262X& DNS \\
2012 & 1672704000 & 10.1017/jfm.2012.257& DNS &2006 & 1186816 & 10.1017/S0022112006002606& DNS \\
2011 & 134217728 & 10.1017/S0022112010005033& DNS &2005 & 251658240 & 10.1017/S0022112005006427& DNS \\
2011 & 2097152 & 10.1017/S0022112010005215& DNS &2005 & 1843200 & 10.1017/S0022112005003964& DNS \\
2011 & 78988950 & 10.1017/S0022112010005082& DNS &2005 & 5376000 & 10.1017/S0022112005003940& DNS \\
2011 & 6291456 & 10.1017/jfm.2011.219& DNS \\
2011 & 26542080 & 10.1017/jfm.2011.252& DNS \\
2011 & 28800000 & 10.1017/S0022112010005902& DNS \\
2011 & 338000000 & 10.1017/S0022112010004866& DNS \\
2011 & 211613535 & 10.1017/S0022112010005094& DNS \\
2011 & 99186753 & 10.1017/jfm.2011.34& DNS \\
2010 & 33600000 & 10.1017/S0022112010001278& DNS \\
2010 & 944000000 & 10.1017/S002211201000039X& DNS \\
2010 & 172800000 & 10.1017/S0022112010000893& DNS \\
2010 & 84756225 & 10.1017/S0022112010003873& DNS \\
2010 & 134217728 & 10.1017/S0022112010000807& DNS \\
2010 & 252593991 & 10.1017/S0022112010001710& DNS \\
2010 & 220000000 & 10.1017/S0022112010000558& DNS \\
\bottomrule
\end{tabular}
\end{table}

\begin{table}\centering
\small
\caption{References used for the data points in Figure 1  (Part 3/3)}
\begin{tabular}{llll|llll}
Year & Max Grid points & DOI & Type & Year & Max Grid points & DOI & Type \\
\midrule
2023 & 570000000 & 10.1017/jfm.2023.175& LES &2016 & 5782500 & 10.1017/jfm.2016.191& LES \\
2023 & 16777216 & 10.1017/jfm.2023.649& LES &2016 & 20480000 & 10.1017/jfm.2016.519& LES \\
2023 & 17039360 & 10.1017/jfm.2022.969& LES &2015 & 149760 & 10.1017/jfm.2015.29& LES \\
2023 & 30000000 & 10.1017/jfm.2023.331& LES &2015 & 4718592 & 10.1017/jfm.2015.604& LES \\
2023 & 12308967 & 10.1017/jfm.2023.575& LES &2015 & 134217728 & 10.1017/jfm.2015.249& LES \\
2023 & 19906560 & 10.1017/jfm.2023.499& LES &2015 & 134217728 & 10.1017/jfm.2015.116& LES \\
2023 & 38000000 & 10.1017/jfm.2023.143& LES &2014 & 452984832 & 10.1017/jfm.2014.381& LES \\
2022 & 184320 & 10.1017/jfm.2021.1156& LES &2014 & 1210056704 & 10.1017/jfm.2014.510& LES \\
2022 & 16777216 & 10.1017/jfm.2021.1150& LES &2014 & 6773760 & 10.1017/jfm.2014.379& LES \\
2022 & 840000000 & 10.1017/jfm.2022.692& LES &2014 & 4579600 & 10.1017/jfm.2014.581& LES \\
2022 & 3145728 & 10.1017/jfm.2022.577& LES &2014 & 40000000 & 10.1017/jfm.2014.505& LES \\
2022 & 1500000000 & 10.1017/jfm.2022.471& LES &2013 & 2097152 & 10.1017/jfm.2013.215& LES \\
2022 & 125000000 & 10.1017/jfm.2022.334& LES &2013 & 19300000 & 10.1017/jfm.2013.292& LES \\
2022 & 356352 & 10.1017/jfm.2022.654& LES &2013 & 10100000 & 10.1017/jfm.2013.135& LES \\
2022 & 31850496 & 10.1017/jfm.2021.1127& LES &2013 & 18874368 & 10.1017/jfm.2013.36& LES \\
2022 & 7200000 & 10.1017/jfm.2022.286& LES &2013 & 73711872 & 10.1017/jfm.2012.513& LES \\
2022 & 4047912960 & 10.1017/jfm.2021.1046& LES &2012 & 198656 & 10.1017/jfm.2012.84& LES \\
2021 & 11052800 & 10.1017/jfm.2021.4& LES &2012 & 16777216 & 10.1017/jfm.2012.115& LES \\
2021 & 4182410 & 10.1017/jfm.2021.332& LES &2012 & 150994944 & 10.1017/jfm.2012.73& LES \\
2021 & 40824000 & 10.1017/jfm.2021.261& LES &2012 & 1436506 & 10.1017/jfm.2011.539& LES \\
2021 & 10600000 & 10.1017/jfm.2021.1002& LES &2012 & 6553600 & 10.1017/jfm.2012.160& LES \\
2021 & 80621568 & 10.1017/jfm.2020.943& LES &2012 & 528384 & 10.1017/jfm.2012.150& LES \\
2021 & 712000000 & 10.1017/jfm.2021.714& LES &2011 & 6356992 & 10.1017/S0022112010005367& LES \\
2021 & 704643072 & 10.1017/jfm.2020.858& LES &2011 & 53248000 & 10.1017/jfm.2011.170& LES \\
2021 & 11291423 & 10.1017/jfm.2021.495& LES &2011 & 2097152 & 10.1017/jfm.2011.137& LES \\
2020 & 83886080 & 10.1017/jfm.2020.512& LES &2011 & 1800000 & 10.1017/S0022112011000450& LES \\
2020 & 134217728 & 10.1017/jfm.2020.622& LES &2011 & 25165824 & 10.1017/jfm.2011.342& LES \\
2020 & 8589934592 & 10.1017/jfm.2020.101& LES &2011 & 33554432 & 10.1017/S002211201000580X& LES \\
2020 & 201719808 & 10.1017/jfm.2020.536& LES &2011 & 25772032 & 10.1017/jfm.2011.130& LES \\
2020 & 80000000 & 10.1017/jfm.2019.711& LES &2010 & 4000000 & 10.1017/S0022112010000017& LES \\
2019 & 22118400 & 10.1017/jfm.2019.649& LES &2010 & 10616832 & 10.1017/S0022112010002995& LES \\
2019 & 16777216 & 10.1017/jfm.2018.910& LES &2010 & 43100000 & 10.1017/S0022112009992965& LES \\
2019 & 26000000 & 10.1017/jfm.2018.949& LES &2010 & 31843449 & 10.1017/S0022112010003927& LES \\
2019 & 27000000 & 10.1017/jfm.2019.80& LES &2010 & 1300000 & 10.1017/S0022112010000686& LES \\
2019 & 31997952 & 10.1017/jfm.2019.481& LES &2010 & 16777216 & 10.1017/S002211200999303X& LES \\
2019 & 134217728 & 10.1017/jfm.2019.591& LES &2009 & 2372500 & 10.1017/S0022112009006739& LES \\
2019 & 209715200 & 10.1017/jfm.2019.360& LES &2009 & 3538944 & 10.1017/S0022112009006867& LES \\
2019 & 294912 & 10.1017/jfm.2018.808& LES &2009 & 9208320 & 10.1017/S0022112008004722& LES \\
2019 & 327680 & 10.1017/jfm.2018.838& LES &2009 & 14515200 & 10.1017/S0022112009007277& LES \\
2018 & 100663296 & 10.1017/jfm.2018.644& LES &2009 & 134534400 & 10.1017/S0022112008005661& LES \\
2018 & 608000000 & 10.1017/jfm.2018.585& LES &2009 & 44346771 & 10.1017/S0022112009005801& LES \\
2018 & 400000000 & 10.1017/jfm.2018.470& LES &2008 & 25000000 & 10.1017/S0022112007009664& LES \\
2018 & 452984832 & 10.1017/jfm.2018.417& LES &2008 & 426951 & 10.1017/S0022112008003443& LES \\
2017 & 76800000 & 10.1017/jfm.2016.841& LES &2008 & 73728 & 10.1017/S0022112008001079& LES \\
2017 & 2147483648 & 10.1017/jfm.2017.172& LES &2007 & 1048576 & 10.1017/S002211200700599X& LES \\
2017 & 181000000 & 10.1017/jfm.2017.20& LES &2007 & 893952 & 10.1017/S0022112006004587& LES \\
2017 & 4500000000 & 10.1017/jfm.2017.187& LES &2007 & 20330271 & 10.1017/S0022112007006842& LES \\
2017 & 3145728 & 10.1017/jfm.2017.450& LES &2007 & 6000000 & 10.1017/S0022112006004502& LES \\
2016 & 2097152 & 10.1017/jfm.2015.744& LES &2007 & 32768 & 10.1017/S0022112007006155& LES \\
2016 & 262852998 & 10.1017/jfm.2016.628& LES &2007 & 11520000 & 10.1017/S002211200700897X& LES \\
2016 & 5782500 & 10.1017/jfm.2016.191& LES &2007 & 7000000 & 10.1017/S0022112006003235& LES \\
2016 & 20480000 & 10.1017/jfm.2016.519& LES &2006 & 18598932 & 10.1017/S0022112006000930& LES \\
2015 & 149760 & 10.1017/jfm.2015.29& LES &2006 & 50855936 & 10.1017/S0022112006009475& LES \\
\bottomrule
\end{tabular}
\end{table}

\bibliographystyle{ieeetr} 
\bibliography{references.bib}

\end{document}